
\let\nopictures=Y
\input tables
\ifx\twocolin Y
        \font\zfont = cmss10
        \else\font\zfont = cmss10 scaled \magstep1 \fi
\def\ZZ{\hbox{\zfont Z\kern-.4emZ}}
\def\RR{\hbox{\zfont F\kern-.4emR}}
\def\bigone{\hbox{1\kern -.23em {\rm l}}}
\ifx\twocolin Y
        \font\smfont = cmr6
        \else\font\smfont = cmr6 scaled \magstep1 \fi
\edef\twoth{\hbox{${\smfont {2\over 3}}$}}
\edef\oneth{\hbox{${\smfont {1\over 3}}$}}
\edef\onesx{\hbox{${\smfont {1\over 6}}$}}
\edef\onehf{\hbox{${\smfont {1\over 2}}$}}

\def\bfphi{\hbox{$\phi\kern-.55em\phi$}}
\def\ls#1{_{\lower1.5pt\hbox{$\scriptstyle #1$}}}

\def\SCIPP{\centerline{\it Santa Cruz Institute for Particle Physics}
\centerline{\it University of California, Santa Cruz, CA 95064}}
\def\bibj#1#2{{\bibit #1~}{\bibbf #2}}
\def\prl{Phys. Rev. Lett.}
\def\prd{Phys. Rev.}
\def\npb{Nucl. Phys.}
\def\plb{Phys. Lett.}
\def\bibit{\it}
\def\bibbf{\bf}

\overfullrule 0pt

\Pubnum{SCIPP 93/27}
\ifx\nopictures Y \titlepage
        \else\treetitle\fi

\title{{ INVESTIGATION OF DISCRETE GAUGE ANOMALIES IN STRING THEORY}\foot{
Work supported in part by the U.S. Department of Energy.}}
\author{D. A. MacIntire\foot{Address after
September 1, 1993: Department of Physics, University of Colorado,
Boulder, CO 80309.}}
\address{\SCIPP}
\ifx\twocolin Y
        \vskip 0.9cm
        \else\vskip 1.30cm\fi

\vbox{
\centerline{\bf Abstract}
In string theory, there are no continuous global symmetries.
Discrete symmetries frequently appear, and these can often be
understood as unbroken subgroups of larger, spontaneously
broken gauge symmetries (discrete gauge symmetries).  In
cases which have been studied previously, anomalies in these
symmetries could always be cancelled by a Green-Schwarz
mechanism.  In the present work, we describe results of an
investigation of a large number of $\ZZ_3$ orbifold models with
two and three Wilson lines.  We find that the discrete
gauge anomalies can always be cancelled.}

\endpage

\REF\banksdixon {T. Banks and L. Dixon, \bibj{\npb}{B307} (1988) 93.}
\REF\gds {L. Krauss and F. Wilczek, \bibj{\prl}{62} (1989) 1221;
L.E. Ib\'a\~nez and G.G. Ross, \bibj{\plb}{260B} (1991) 291;
\bibj{\npb}{B368} (1992) 3; J. Preskill, S. Trivedi,
F. Wilczek and M. Wise, \bibj{\npb}{B363} (1991) 207.}
\REF\banksdine{T. Banks and M. Dine, \bibj{\prd}{D45} (1992) 424.}
\REF\dlm{M. Dine, R. Leigh, D.A. MacIntire
(unpublished).}

In the absence of a definitive understanding of the vacuum structure
of string theory, it is of interest to enumerate those properties
of the theory that may be considered generic or universal. One
well known example\refmark{\banksdixon} is that all continuous
symmetries in string theory are gauge symmetries. This remarkably
powerful conclusion is reached by noting that for every current
there is a corresponding gauge boson vertex operator. Whether or
not this conclusion can be extended to {\it discrete} symmetries is
of great interest. In field theory, gauged discrete
symmetries,\refmark{\gds} by definition, arise through the spontaneous
breakdown of continuous local symmetries. It is clear then that any
such discrete symmetry is anomaly-free, this being inherited from
the underlying symmetry.

At low energies, this anomaly freedom leads to a useful constraint
obtained by requiring that the 't Hooft effective
fermion vertex be invariant under the discrete symmetry. Otherwise,
instanton effects would lead to explicit violations of the symmetry,
and it could not be gauged. In string theory, previous studies
suggest that a slightly more general result holds.\refmark{\banksdine}
In this case the instanton vertex is accompanied by an exponential of
the model-independent axion field (the partner of the dilaton, whose
vacuum expectation value sets the value of the gauge coupling).
A discrete symmetry can be gauged only if the
transformation of the fermionic part of the instanton vertex can be
cancelled by assigning a shift to the model-independent axion. This is
just the four-dimensional analogue of the Green-Schwarz mechanism
applied to discrete symmetries.  In a large number of Calabi-Yau
compactifications (with or without
Wilson lines), it has been shown that such
an assignment is possible.\refmark{\dlm}
We should note, however, that
there is no deep understanding of how this transformation law
arises.  (Nor can one be certain that it {\it does} arise.)
This is to be contrasted with the Green-Schwarz terms
for continuous symmetries, where one can check that the appropriate
couplings (to gauge bosons) appear in perturbation theory.

This cancellation is non-trivial in the case in which there is more
than one non-Abelian gauge group. The discrete anomalies of each
gauge group must match in order that they all be cancelled by the
transformation of the one model-independent axion. If all discrete
symmetries are gauge symmetries in string theory, then they must pass
this rather stringent test.

In this paper we report the results of an investigation of discrete
gauge anomalies in a large number of models orbifold models with Wilson lines.
We find that in all models investigated, the discrete gauge anomalies can
be cancelled through a transformation of the axion.

\REF\zorb {L. Dixon, J.A. Harvey, C. Vafa and E. Witten, \bibj{\npb}{B261}
(1985) 678; \bibj{\npb}{B274} (1986) 285;
L.E. Ib\'a\~nez, H.P. Nilles, F. Quevedo, \bibj{\plb}{187B} (1987) 25.}

\chapter{$\ZZ_3$ Orbifolds with Wilson Lines}

The $\ZZ_3$ orbifold compactification of the heterotic string is simple
and well known.\refmark{\zorb} We begin by reviewing the relevant formalism.

The $\ZZ_3$ orbifold with Wilson lines is obtained by modding $\RR^6$ by
the space group $S$, consisting of rotations in three two-dimensional planes
and translations:
$$ S\, :\, Z \rightarrow \Theta Z + {\rm v}.\eqn\szthz$$
Associated with this is a transformation of the $E_8\times E_8$ lattice
$$\eqalign{\left(\Theta,0\right) \rightarrow \left(\bigone,{\rm V}^I\right) \cr
\left(\bigone,e^i_m\right)\rightarrow\left(\bigone,{\rm a}^I_m\right)
}\eqn\trans$$
and a transformation of the bosonized fermion lattice:
$$\eqalign{\left(\Theta,0\right) \rightarrow \left(\bigone,
{\rm v}\ls{\scriptscriptstyle R}\right) \cr
\left(\bigone,e^i_m\right)\rightarrow\left(\bigone,0\right) .
}\eqn\transpr$$

These models are completely specified by the shift vector
${\bf V}$ and Wilson lines ${\rm a}_m^I, m=1,3,5$.
These vectors must satisfy the modular invariance constraints
$$3\,\left({\bf V}+\sum_m n^m {\bf a}_m\right)^2=0\bmod 2,
\quad n^m=0,\pm1\eqn\modinv$$
and the requirements $3 {\bf V}, 3 {\bf a}_m\in E_8\times E_8$.
The gauge bosons for these models correspond to vectors
${\bf P}\in E_8\times E_8$ satisfying
$${\bf P}^2=2, \quad {\bf P\cdot V}\in\ZZ
,\quad {\bf P\cdot a}_m\in\ZZ,\,\,\forall m .\eqn\ggbos$$
Massless matter states from the untwisted sector are given by
$${\bf P}^2=2, \quad {\bf P\cdot V}={1\over3}\bmod 1,\quad {\bf P\cdot
a}_m\in\ZZ,\,\,\forall m.\eqn\matst$$
Massless states from the twisted sectors are given by
$$ \left({\bf P}+{\bf V}+\sum_m n^m {\bf a}_m\right)^2=
{4\over 3} \quad {\rm (multiplicity\, 1)}\eqn\twstone$$
and
$$ \left({\bf P}+{\bf V}+\sum_m n^m {\bf a}_m\right)^2=
{2\over 3} \quad {\rm (oscillator\, states, \, multiplicity\,
3).}\eqn\twsttwo$$
With three inequivalent non-zero Wilson lines there are twenty-seven
distinct twisted sectors.

Whereas the modding $\Theta$ is chosen to be the diagonal $\ZZ_3$ given
by $\Theta^i_j = \delta^i_j \exp(2\pi i \eta^j/3 )$ with $\eta=(1,1,-2)$,
we will look for anomalies in the discrete $\ZZ_3$ rotation of {\it one}
of the three $SU(3)$ lattices, referred to here as $G_1$. Before modding,
it is clear that this
is just a general coordinate transformation in the internal space. That
it remains a symmetry of the interacting theory may be verified by
checking that it commutes with the BRST operator. To determine the
action of this discrete rotation on the fermions of the model, it
is simplest to bosonize the right-moving fermions.  In the light-cone
gauge, the quantized momenta of bosons describing the (GSO projected)
Ramond sector belong to the $(s)$ representation of $SO(8)$.  The positive
helicity momenta are therefore of the form
$${\bf k}=\left(n_1+\onehf,n_2+\onehf,n_3+\onehf;+\onehf\right),\qquad
\sum_i n_i=0\bmod 2, \qquad n_i\in\ZZ\eqn\Rstates$$
where the last component of $\bf k$ is the spacetime helicity.

A $G_1$-rotation of $2\pi/3$ of the first $SU(3)$ lattice
corresponds to a shift of the bosonized field:
$${\bfphi}\rightarrow {\bfphi}+2\pi{\bf v}_1\qquad {\bf v}_1
=\left(\oneth,0,0;0\right).\eqn\phishift$$
The right-moving component of a state will therefore acquire a phase
under the discrete $G_1$-rotation given by $e^{2\pi i {\bf k\cdot v}_1}.$
We see also from this consideration that $G_1$ is an R-symmetry;
the gauginos (and the supercharges),
with ${\bf k} = \left({1\over2},{1\over2}, {1\over2};{1\over2}\right)$,
transform under $G_1$ by a phase $e^{i\pi /3}$.

In the twisted sectors, the momenta \Rstates\ are shifted according to
$${\bf k}\rightarrow {\bf k}+{\bf v_{\scriptscriptstyle R}}\qquad
{\bf v_{\scriptscriptstyle R}}=
\left(\oneth,\oneth,-\twoth;0\right).\eqn\twshift$$
Massless twisted (matter) sector states must have a right-moving component with
momentum given by\foot{We are indebted to Luis Ibanez for pointing out that one
must be careful with this helicity assignment for the twisted sector.}
$${\bf k}+{\bf v_{\scriptscriptstyle R}}=\left(\onesx,\onesx,
\onesx;-\onehf\right).\eqn\twmom$$
These states acquire a phase under the discrete rotation given by
$e^{\left\{2\pi i ({\bf k+v_{\scriptscriptstyle R})\cdot v}_1\right\}}.$

$G_1$ will also have an effect on the left-moving component of the
massless oscillator states of the form
$\ket{\bf k}_R\tilde\alpha^i_{-1/3}\ket{\bf P}_L,\, i=1,2,3.$
Because $\tilde\alpha^i$ (in particular, $\tilde\alpha^1$)
transforms like $Z^i$ under the discrete rotation, there is an
additional phase of $e^{2\pi i/3}$ that must be take into account for
states of this type.

\chapter{Models}

The 't Hooft vertex for a given model and non-Abelian gauge group is a product
of the form
$$e^{ia/f_a}\prod_{massless} \psi\psi \cdots \psi$$
with one field $\psi$ for each zero mode of a given multiplet.  There is
a single zero mode for each doublet of $SU(2)$ or triplet of $SU(3)$,
and four or six zero modes for the $SU(2)$ or $SU(3)$ gauginos respectively.
Defining $\alpha=e^{2\pi i/3}$ and using the above relations, we find that
for a model with $N_{UT}$ untwisted matter doublets or triplets, and $N_{TW}$
matter  doublets or triplets, the transformation
of the 't Hooft vertex under the discrete transformation is given by
$$e^{ia/f_a}\prod_{massless} \psi\psi \cdots \psi\rightarrow
\alpha^{-N_{UT}/6} \alpha^{N_{TW}/6} \alpha^{C_G}
e^{ia/f_a}\prod_{massless} \psi\psi\cdots\psi\eqn\tHoofttrans$$
with $C_G$ the quadratic Casimir.
If the phase given in Eq. \tHoofttrans\ is non-zero, the discrete symmetry
is anomalous. However, note the presence of the model-independent axion
in the 't Hooft vertex; its form is universal for each non-Abelian group.
The discrete anomaly may be cancelled by assigning a $G_1$-transformation
to the axion if and only if the transformation \tHoofttrans\ is the
{\it same} for each non-Abelian group. In earlier studies, it was
shown that many models do satisfy this condition.  For example,
if one takes the Calabi-Yau manifold $Y^4_5$, with Wilson lines,
one finds that this condition is satisfied for any choice of
Wilson line, for either the $E_8 \times E_8$
or $O(32)$ theories.

An example of a particular model with cancellable discrete gauge anomalies
is generated by the vectors:
$$\eqalign{{\bf V}&=\left( -\twoth,0,-\twoth,0,0,-\twoth,0,0\right)
                    \left(0,0,-\twoth,\twoth,\twoth,0,0,0\right)\cr
{\bf a_1}&=\left(0,0,0,\twoth,0,0,0,0\right)
           \left(0,\oneth,0,0,0,0,-\oneth,0\right)\cr
{\bf a_3}&=\left(0,0,0,-\oneth,\twoth,\oneth,\twoth,-\twoth\right)
           \left(-\oneth,\twoth,\twoth,0,0,0,0,-\oneth\right)\cr
{\bf a_5}&=\left(0,0,0,0,0,\twoth,0,0\right)
           \left(0,0,-\oneth,0,\oneth,0,0,0\right).\cr}\eqn\willines$$
This model has the gauge group $SU(3)\otimes SU(2)\otimes SU(2)$ with
$SU(2)$ charge operators given by:
$$\eqalign{I_3&=\onehf (1,0,-1,0,0,0,0,0)(0,0,0,0,0,0,0,0)\cr
          I'_3&=\onehf (0,0,0,0,0,0,0,0)(1,0,0,0,0,0,0,-1)\cr}$$
The full spectrum of this model is
$$({\bf 3},{\bf 1},{\bf 1})+({\bf \bar 3},{\bf 1},{\bf 1})
+20\left[ ({\bf 1},{\bf 2},{\bf 1})+({\bf 1},{\bf 1},{\bf 2})\right]
+ 205 ({\bf 1},{\bf 1},{\bf 1}).
\eqn\spectr$$
Six of the twenty $SU(2)$ doublets are from the untwisted sector, and the
remaining doublets are from the twisted sector, for both $SU(2)$'s.
For the $SU(3)$, both the $\bf 3$ and $\bf \bar 3$ are twisted matter.
The discrete transformation for the fermionic part of the 't Hooft action
for each of the gauge groups is $\alpha^{1/3}$.

On the order of $10^5$ randomly generated $\ZZ_3$ orbifold models
were investigated.  The search included models with two or three
non-zero Wilson lines.  We found that in all cases the discrete
anomalies can be cancelled through a transformation of the model independent
axion.

\chapter{Conclusions}
Previously, only a limited number of string theory models have been checked
for the cancellation of discrete gauge anomalies.  In this paper we report
on an investigation of a huge number of models, and have found that this
cancellation always occurs.
While searches to date have involved only particular types of
models, and the search should be expanded to other types of string theory
models, we suspect that there exists some general connection between the
requirement of modular invariance and the cancellation of
discrete gauge anomalies.

\endpage
\refout
\end